\documentstyle[preprint,aps,graphicx]{revtex}

\def\red{
}
\def\black{
}

\def\URLtilde{\lower0.2em\hbox{$\tilde{\phantom{a}}$}}
\def\mycomm#1{\hfill\break\strut\kern-3em{\red\tt ====> #1\black}\hfill\break}
\def\mycommNL#1{\strut\kern-3em{\tt ====> #1}\hfill\break}

\catcode`\@=11 % This allows us to modify PLAIN macros.
\def\lsim{\mathrel{\mathpalette\@versim<}}
\def\gsim{\mathrel{\mathpalette\@versim>}}
\def\@versim#1#2{\vcenter{\offinterlineskip
        \ialign{$\m@th#1\hfil##\hfil$\crcr#2\crcr\sim\crcr } }}
\catcode`\@=12 % at signs are no longer letters

\def\sbar{\hbox{$\bar s$}}

\def\thetap{\hbox{$\Theta^+$}}

\def\eqref#1{(\ref{#1})}

\makeatletter
\def\hlinewd#1{\noalign{\ifnum0=`}\fi
\hrule \@height #1 \futurelet \reserved@a\@xhline}
\def\hwhiteline{\noalign
{\ifnum0=`}\fi\hrule
\@height 0pt\vskip 1.0ex\futurelet \reserved@a\@xhline}
\makeatother
\def\gray{\special{ps: 0.40 setgray}}
\def\black{\special{ps: 0.0 setgray}}

\newcommand{\mydraft}{
\newcount\timecount
\newcount\hours \newcount\minutes  \newcount\temp \newcount\pmhours

\hours = \time
\divide\hours by 60
\temp = \hours
\multiply\temp by 60
\minutes = \time
\advance\minutes by -\temp
\def\hour{\the\hours}
\def\minute{\ifnum\minutes<10 0\the\minutes
    \else\the\minutes\fi}
\def\clock{
\ifnum\hours=0 12:\minute\ AM
\else\ifnum\hours<12 \hour:\minute\ AM
\else\ifnum\hours=12 12:\minute\ PM
    \else\ifnum\hours>12
     \pmhours=\hours
     \advance\pmhours by -12
     \the\pmhours:\minute\ PM
     \fi
    \fi
\fi
\fi
}
\def\fullclock{\hour:\minute}
\begin{centering}
\gray
\font\Hugett  =cmtt12 scaled\magstep4
\hbox{\Hugett Draft:\today,\clock}
\black
\end{centering}
\vskip -1.7cm
$\phantom{a}$
} % end of \draft definition
\def\beq#1{\begin{equation} \label{#1}}
\def\eeq{\end{equation}}

\newskip\humongous \humongous=0pt plus 1000pt minus 1000pt

\newif\ifdtup

\begin{document}
{\tighten
 \preprint {\vbox{
  \hbox{$\phantom{aaa}$}
  \vskip-0.5cm
%\hbox{\today}
%\hbox{}
\hbox{Cavendish-HEP-04/15}
\hbox{TAUP 2769-04}
\hbox{WIS/11/04-May-DPP}
\hbox{ANL-HEP-PR-04-45}
}}

\title{Why the $\Theta^+$ is seen in some experiments and not in others
-- \\ a possible explanation}

\author{Marek Karliner\,$^{a,b}$\thanks{e-mail: \tt marek@proton.tau.ac.il}
\\
and
\\
Harry J. Lipkin\,$^{b,c}$\thanks{e-mail: \tt
ftlipkin@clever.weizmann.ac.il} }
\address{ \vbox{\vskip 0.truecm}
$^a\;$Cavendish Laboratory\\
Cambridge University, England;\\
%\mbox{}\\
and\\
$^b\;$School of Physics and Astronomy \\
Raymond and Beverly Sackler Faculty of Exact Sciences \\
Tel Aviv University, Tel Aviv, Israel\\
\vbox{\vskip 0.0truecm}
$^c\;$Department of Particle Physics \\
Weizmann Institute of Science, Rehovot 76100, Israel \\
and\\
High Energy Physics Division, Argonne National Laboratory \\
Argonne, IL 60439-4815, USA\\
}
\maketitle
%\mydraft
\begin{abstract}%

The contradiction between evidence for and against the existence
of the $\Theta^+$ pentaquark might be resolved if it only appears
as a result of a particular production mechanism  which is present
in some experiments and absent in others. We examine the
implications of $\Theta^+$ production via decay of a cryptoexotic
$N^*$ resonance with a mass of about 2.4 GeV corresponding to a
peak in the experimental data for the invariant mass of the
($\Theta^+,K^-$) system. Further experimental checks are
suggested.

\end{abstract}% %}
% end tighten

\vfill\eject

The recent experimental discovery
\cite{Nakano:2003qx} and subsequent confirmation
\cite{Barmin:2003vv}-\kern-0.5em\cite{Troyan:2004wp}
of an exotic 5-quark $K N$ resonance \thetap\
with \hbox{$S={+}1$}, a mass of $\sim$1540 MeV,
a very small width $\lsim 20$
MeV (possibly as little as $1{\div2}$~MeV~\cite{ThetaWidth}),
and a presumed quark configuration $uudd\sbar$  has given rise
to a number of experiments with contrary results.
Some experiments see the $\Theta^+$
\cite{Nakano:2003qx}-\kern-0.5em\cite{Troyan:2004wp},
others definitely do not
\cite{BES}-\kern-0.5em\cite{DELPHI}
and give upper limits on its production.

This contradiction is expected to become even sharper as the experiments which
see the $\Theta^+$ have better statistics and rule out the explanation that it
is a statistical fluctuation.  At this point it seems crucial to analyze and
extend both the positive and negative experiments to either establish the
$\Theta^+$ as a real particle and understand this contradiction or to find good
credible reasons against its existence.

Many detailed theoretical pentaquark models have been proposed, but none
address the problem of why certain experiments see it and others do not.
We therefore do not consider them here and refer the reader to the
comprehensive review by Jennings and Maltman\cite{jenmalt}.

Our purpose here is to analyze the
puzzle, suggest one possible explanation and suggest experimental checks.

One possible resolution of this contradiction is that a specific production
mechanism is present in the experiments that see the $\Theta^+$ and is absent
in those that do not see it. The data presented in the CLAS paper
on the reaction $\gamma p \rightarrow \pi^+ K^- K^+ n$
\cite{Kubarovsky:2003fi}, and in
particular the $(K^+K^-n)$ mass distribution in Fig.~5
which shows a peak at the mass of
2.4 GeV suggest
\cite{BurkertNSTAR2004}
that there might be a cryptoexotic $N^*$ resonance with
hidden strangeness.
Searches for such baryon resonances with hidden
strangeness \cite{Landsberg:1999wn}
have indicated possible candidates. Further evidence for this resonance is
hinted at in the preliminary results from NA49 \cite{NA49Theta}.

A cryptoexotic $N^*(2400)$ with hidden strangeness has a mass too 
high to be the
$N^*$ in the same $SU(3)$ multiplet as the $\Theta^+$. It fits naturally
into the $P$-wave $(ud)$ diquark-$ud \bar s$ triquark model
\cite{OlPenta,NewPenta} for the
$\Theta^+$, as an orbital excitation of the $udd\bar s s$ $N^*$ 
in the same $\overline{\hbox{\bf 10}}$. It contains 
a $(ds)$ diquark in the same flavor $SU(3)$ 
multiplet as the $(ud)$ diquark in the $\Theta^+$. Such a $(ds)$ diquark
in a $D$-wave with the $ud \bar s$ triquark
would have a dominant decay into $K^- \Theta^+$ via the diquark transition
$ds \rightarrow ud + K^-$. Decays into a kaon and a hyperon would be
suppressed by the centrifugal barrier forbidding a quark in the triquark
from joining the diquark.

We wish to point out some experimental implications of this possibility and
suggest ways of using experimental data to check whether this  can indeed solve
the puzzle of the contradiction between positive and negative evidence for the
$\Theta^+$.

\begin{enumerate}
 \item
 All experiments which see the $\Theta^+$ and have sufficient energy for
producing the $N^*(2400)$ should look for an accompanying $K^-$ or $K_s$ and
examine the mass spectrum of the $K^-\Theta^+$ and $K_s \Theta^+$
systems.\footnote{In some photoproduction experiments, e.g. SPRING-8 and the
lower-energy CLAS-I, the photon energy is too low for this. We thank Danny
Ashery for pointing this out. Excitation at such lower energies may be possible
at the lower energy tail of a Breit-Wigner resonance 200 MeV wide or 
using Fermi momentum in experiments on nuclear targets; e.g. $\gamma {}^{12}C$
at SPRING-8, where $E_{CM}(\gamma N)$=2.3 GeV.}

\item There are many rumors and conference presentations
\cite{ALEPH}-\kern-0.5em\cite{DELPHI} about experiments that
searched for pentaquarks and did not find them. These experiments
should not be left in the rumor/slides stage but put on the record
with a careful analysis showing whether they should have been seen
with given specific production mechanisms.

All experiments which did not see the $\Theta^+$ should check whether their
experiment would produce a $K^-\Theta^+$ or $K_s \Theta^+$ resonance in the 2.4
GeV region and whether their analysis would emphasize this region in their
search for the $\Theta^+$. For example, the $B$-decay modes that have been
suggested for pentaquark searches\cite{rosner,Browder:2004mp}
 would not produce this 2.4 GeV $N^*$.
Similar considerations should be applied
to searches in $e^+e^-$ and $\gamma\gamma$ like those proposed in
Ref.~\cite{Armstrong:2003zc}.

\item
The angular distribution of the kaon emitted with the $\Theta^+$ in the
photoproduction reaction
$\gamma p \rightarrow \bar K^o \Theta^+$ for which preliminary data have
recently been presented by CLAS \cite{CLAS-Trento}
carries interesting information.
If it is produced from a cryptoexotic $N^*$, there should be no
forward-backward asymmetry in the kaon angular distribution.
If it is peaked forward, this is meson exchange.
It it is peaked backward, it is baryon exchange
(see related discussion in Ref.~\cite{Karliner:2004qw}).
In this case the same baryon exchange should be seen
in \ $\gamma n \rightarrow K^-  \Theta^+$.
The $\Theta^+$ should be produced equally by photons on protons and neutrons.

\item
The angular distributions in the
photoproduction reaction \ $\gamma p \rightarrow \pi^+ K^- K^+ n$
\ \cite{Kubarovsky:2003fi}
are more complicated, but may still carry interesting information. We
consider two possible production mechanisms for the additional pion.

If the $\pi^+ K^-$ system comes from a $\bar K^*$ resonance, all the above
discussion for the photoproduction reaction $\gamma p \rightarrow \bar K^o
\Theta^+$ applies to the angular distribution of the $\bar K^*$. Models
like Ref.~\cite{Karliner:2004qw})  which explain the narrow width of the
$\Theta^+$ by a suppressed $N K \Theta^+$ coupling relative to $N K^*
\Theta^+$ can be tested here via their prediction that $\Theta^+$
production with a backward $K^*$ should be stronger than the production
with a backward kaon.  Unfortunately measurement of forward-backward
asymmetry is complicated by the presence of a strong forward-peaked
background due to $t$-channel exchange that is most simply treated by
cutting out all forward-peaked events including the signal.

If the reaction goes via the cryptoexotic $N^*$ and is described by the
diagram 3a of Ref.~\cite{Kubarovsky:2003fi}, the pion goes forward and
everything else is in the target fragmentation region.
The latter possibility is strengthened by the fact
that the $\pi^- p$ cross section data have a gap
in the mass range $2.3-2.43$ GeV
\cite{BurkertNSTAR2004}.

\item

The production of $\Theta^+$ by baryon exchange is related to reactions
between normal nonexotic hadrons that can go by exchange of an exotic
positive-strangeness baryon. The baryon exchange diagram proposed in
\cite{CLAS-Trento} for $\Theta^+$ photoproduction with an outgoing kaon is
simply related to the  backward $K^-p$ charge-exchange diagram shown in Fig. 1
of \cite{Karliner:2004qw}. The lower $K N \Theta^+$ vertices are the same; the
upper vertex is also $K N \Theta^+$ for $K^-p$ charge-exchange but is  $\gamma
\Theta^+ \Theta^+$ for $\Theta^+$ photoproduction.

If this diagram contributes appreciably to $\Theta^+$ photoproduction,  it
indicates that the contribution of the  $K N \Theta^+$ vertex is appreciable
and should also contribute appreciably to backward $K^- p$ charge-exchange.
There may even be some backward $K^- p$ charge-exchange data available
previously ignored, because  everyone knew that there were no positive
strangeness baryons to produce this baryon exchange.

\item
The $\Theta^+$ is a baryon containing a strange antiquark.
In the low-energy
photoproduction experiments these constituents are already present in the
initial state, the baryon in the target and the strange antiquark in the
strange component of the photon, which is known. In other experiments where
baryon number and strangeness must be created from gluons, the cost of
baryon antibaryon and strangeness-antistrangeness production by gluons
must be used to normalize the production cross section in comparison with
the photoproduction cross sections. This can be done experimentally by
measuring the baryon-antibaryon production and strange pair production in
the same experiment that does not see the $\Theta^+$.

One can also tune this kind of estimates by comparing the rate of
anti deuteron and antiproton production in a given experiment.
Such an analysis has been carried out by H1  \cite{Aktas:2004pq},
yielding antideuteron/antiproton ratio \ $\bar d/\bar p = 5.0 \pm
1.0 \pm 0.5\times10^{-4}$.

On the other hand, although LEP experiments produce roughly one
proton per $Z^0$ decay \cite{Knowles:1997dk} and have accumulated
millions of $Z^0$ decays on tape, very little is known about
antideuteron production at LEP. The one theoretical prediction we
are aware of is Ref.~\cite{Gustafson:1993mm}, which uses the Lund
string fragmentation model to predict  $5 \times 10^{-5}$
deuterons per $Z^0$ decay. The only relevant experimental
publication we are aware of is from OPAL \cite{Akers:1995az},
which reports exactly {\em one} antideuteron candiate event which
was eventually dismissed because it did not pass through the
primary vertex. From this OPAL infers at 90\% confidence level an
upper limit on antideuteron production of $0.8 \times 10^{-5}$
anti-deuterons per $Z^0$ in the momentum range $0.35 < p < 1.1$
GeV.

A recent estimate \cite{SloanPC} based on this data concludes that
$\bar d/\bar p < 1.6 \times 10^{-4}$ which is significantly less
than the ratio reported by H1  \cite{Aktas:2004pq}.\footnote{We
thank T.~Sloan for discussion of this point.} The reason for this
presumed difference is unknown at present.
 It would be very
valuable to have more information on antideuterons from the LEP
experiments.

\item ZEUS has observed both $\Theta^+$ and its antiparticle,
$\bar \Theta^-$ \cite{Chekanov:2004kn}. It is important for ZEUS
to provide information about the relative number of
anti-$\Theta$-s and the number of antiprotons. This would give the
probability of creating a $\Theta^+$ when the baryon is already
present. This probability has to be folded into any experiment
(e.g. at LEP) which does not have an initial baryon, does not see
the $\Theta^+$, and wants to interpret their upper limit as
significant evidence against it. We note in passing that
a statement from H1 regarding the $\Theta^+$ is expected in near future.

\item The cryptoexotic $N^*$ would be expected to have other decay modes.
In the diquark-triquark model the dominant other decay mode is the SU(3)
partner of the $K^- \Theta^+$ decay giving a pion and a P-wave nonstrange
pentaquark with hidden strangeness. Decays into a strange meson carrying
the strange antiquark and a normal baryon; e.g. $K \Lambda$, $K \Sigma$,
$K\Sigma^*$, $\phi N$, are suppressed by the centrifugal barrier in the
D-wave diquark-triquark model but may be appreciable in other models.
Searching for these other decay modes would would give further evidence
for this cryptoexotic resonance and this model for pentaquark production.
The relative branching ratios would also provide information about the
structure of this $N^*$. The $N^*$ is an isospin doublet and both charge
states $N^{*+}$ and $N^{*o}$ should be observed.

\item The cryptoexotic $N^*$ with hidden strangeness could have a
partner $N^*_{\bar c c}$ with hidden charm, obtained by replacing
the $s \bar s$ pair by a $c \bar c$ pair. This would then be
observable as a $D \Theta_c$ or $D^* \Theta_c$ resonance seen as a
$D \bar D N$, $D^* \bar D^* N$, $D^* \bar D N$ or $\bar D^* D N$
narrow resonance near the mass of $2.4 + 2 [M(\Lambda_c) -
M(\Lambda)] \sim 4.7$ GeV.\footnote{We thank Uri Karshon for
discussion on the interplay of  $N^*_{\bar c c}$ mass estimate vs.
the relevant thresholds.} In any model with orbital excitation the higher
mass of the $c \bar c$ pair will reduce the kinetic energy. A~quantitative
estimate of this reduction is highly model dependent.

\end{enumerate}

If the $\Theta^+$ is a positive parity pentaquark, as suggested e.g.
in correlated quark models
\cite{OlPenta,NewPenta}, \cite{JW}, \cite{jenmalt},
 there must be a $P$-wave orbital excitation that leads to two states
 having $J=1/2$
and $J=3/2$ with a small spin-orbit splitting\cite{clodudek}  of
the order of 50 MeV.   Both states would be expected to be
produced roughly equally in the $K \Theta^+ $ decay of a higher $N^*$
resonance with the same orbital partial wave, except for the case
where the $N^*$ has $J^P = (1/2)^-$. The more complicated angular
distributions from the production and decay of the $J=3/2$ state
can provide additional information.

The discussion of possible $J=3/2$ partners is especially relevant
in view of
 a recent preliminary report from CLAS
\cite{CLAS-Trento} indicating a possible existence of two peaks in
the $K^+ n$ invariant mass -- at $1523\pm5$ and $1573\pm5$ MeV,
with estimated statistical significance of 4$\sigma$ and
6$\sigma$, respectively. It is very important that other
experiments check this observation.

The preceding discussion focused on the $\Theta^+$, but some of
the above comments apply also to the searches for the
$\Xi^{--}$,
$\Theta_c$, $\Theta_b^+$ and  other pentaquarks.

If the $\Theta^+$ is confirmed, the likelihood that other members
of the antidecuplet exists is quite high \cite{DPP},
\cite{OlPenta,NewPenta}, \cite{JW} and possibly there are
additional exotic multiplets whose properties can be inferred from
those of the $\Theta^+$, see e.g.
\cite{Kopel}-\kern-0.3em\cite{Ellis:2004uz}.

So far, one published experiment reported observing the
$\Xi^{--}$, i.e. the $ddss\bar u$ pentaquark \cite{NA49} at
$1.862\pm0.002$ GeV and width below the detector resolution of
about 18 MeV, as well as a candidate at the same mass for the
$\Xi_{3/2}^o$ member of the corresponding $I=3/2$ isomultiplet,
with quark content $uss \bar q q$, where $q=u,d$. A critical
discussion of the NA49 results appears in
Ref.~\cite{Fischer:2004qb}. There are conference talks from WA89
\cite{WA89}, CDF \cite{CDF} and ZEUS~\cite{ChekanovDIS2004},
reporting null search results, but again no papers.

 The mass of the $\Xi^{--}$ as reported by NA49
\cite{NA49} seems rather high compared with the theoretical
expectations \cite{OlPenta,NewPenta}, \cite{JW} based on the
$\Theta^+$ mass. Moreover, recently we derived an upper bound on
the mass difference between the $\Xi^{--}$ and $\Theta^+$
\cite{varxistar}. This bound is more than 20 MeV below the
experimentally reported $\Xi^{--}-\Theta^+$ mass difference.

The existence of $\Theta^+$ would also make it very likely that
its anti-charmed and anti-bottom relatives $\Theta_c$ and
$\Theta_b^+$ exist. Theoretical predictions based on the presumed
quark structure of the $\Theta^+$ place the $\Theta_c$ mass
between 3 GeV \cite{Karliner:2003si} and 2.7 GeV \cite{JW} and the
$\Theta_b^+$ mass between 6.40 GeV \cite{Karliner:2003si} and 6.05
GeV \cite{JW}, where the lower values are below threshold for
strong decays.

Recently the H1 Collaboration reported evidence for
a narrow anti-charmed baryon state,
a resonance in $D^{*-} p$ and $D^{*+}\bar p$
with a mass of $3099 \pm 3 \pm 5$ MeV and a measured Gaussian width of
$12 \pm 3$  MeV  \cite{H1_Thetac}
A parallel analysis by ZEUS sees no signal \cite{ZEUS_Thetac}.
ALEPH has also reported a null result at a conference
\cite{ALEPH} and FOCUS announced null search results on a Web page
\cite{FOCUS}.
Again, we can only stress again
the importance of having these results written up.

\section*{Note added}
After this work appeared in the arXiv, Ref.~\cite{AS}
pointed out additonal tentative evidence 
for $N^*$ with hidden strangeness and a mass around 2400 MeV
\cite{Landsberg:1999wn}.

\section*{Acknowledgements}

The research of one of us (M.K.) was supported in part by a grant from the
United States-Israel Binational Science Foundation (BSF), Jerusalem.
The research of one of us (H.J.L.) was supported in part by the U.S. Department
of Energy, Division of High Energy Physics, Contract W-31-109-ENG-38.

We thank
Danny Ashery,
Daniel Barna,
Stan Brodsky,
Volker Burkert,
Frank Close,
Karin Daum,
Sergei Chekanov,
Leonid Gladilin,
Ken Hicks,
Kreso Kadija,
Uri Karshon,
Terry Sloan,
Igor Strakovsky
and
Bryan Webber
for discussions.

%----------------------------------------------------------------------
% This prevents REFERENCES from forcing a page break
%\def\newpage{\vskip10ex}
%
\catcode`\@=11 % This allows us to modify PLAIN macros
\def\references{
\ifpreprintsty \vskip 10ex
%\ifpreprintsty \newpage
%
\hbox to\hsize{\hss \large \refname \hss }\else
\vskip 24pt \hrule width\hsize \relax \vskip 1.6cm \fi \list
{\@biblabel {\arabic {enumiv}}}
{\labelwidth \WidestRefLabelThusFar \labelsep 4pt \leftmargin \labelwidth
\advance \leftmargin \labelsep \ifdim \baselinestretch pt>1 pt
\parsep 4pt\relax \else \parsep 0pt\relax \fi \itemsep \parsep \usecounter
{enumiv}\let \p@enumiv \@empty \def \theenumiv {\arabic {enumiv}}}
\let \newblock \relax \sloppy
 \clubpenalty 4000\widowpenalty 4000 \sfcode `\.=1000\relax \ifpreprintsty
\else \small \fi}
\catcode`\@=12 % at signs are no longer letters
%-----------------------------------------------------------------
%{\tighten

} % end of global \tighten

\begin{thebibliography}{99}

%\cite{Nakano:2003qx}
\bibitem{Nakano:2003qx}
T.~Nakano {\it et al.}  [LEPS Coll.],
%``Observation of S = +1 baryon resonance in photo-production from
%neutron,''
Phys.\ Rev.\ Lett.\  {\bf 91}, 012002 (2003), hep-ex/0301020.
%%CITATION = HEP-EX 0301020;%%

\bibitem{Barmin:2003vv}
V.~V.~Barmin {\it et al.}  [DIANA Coll.],
%``Observation of a baryon resonance with positive strangeness in K+
%collisions
%with Xe nuclei,''
Phys.\ Atom.\ Nucl.\  {\bf 66}, 1715 (2003)
[Yad.\ Fiz.\  {\bf 66}, 1763 (2003)], hep-ex/0304040.
%%CITATION = HEP-EX 0304040;%%

%\cite{Stepanyan:2003qr}
\bibitem{Stepanyan:2003qr}
S.~Stepanyan {\it et al.}  [CLAS Coll.],
%``Observation of an exotic S = +1 baryon in exclusive photoproduction from
%the deuteron,''
Phys.\ Rev.\ Lett.\  {\bf 91}, 252001 (2003), hep-ex/0307018.
%%CITATION = HEP-EX 0307018;%%

%\cite{Barth:2003es}
\bibitem{Barth:2003es}
J.~Barth {\it et al.}  [SAPHIR Coll.],
%``Observation of the positive-strangeness pentaquark Theta+ in
%photoproduction
%with the SAPHIR detector at ELSA,''
hep-ex/0307083.
%%CITATION = HEP-EX 0307083;%%

%\cite{Asratyan:2003cb}
\bibitem{Asratyan:2003cb}
A.~E.~Asratyan, A.~G.~Dolgolenko and M.~A.~Kubantsev,
%``Evidence for formation of a narrow K0(S)p resonance with mass near
%1533-MeV in neutrino interactions,''
hep-ex/0309042.
%%CITATION = HEP-EX 0309042;%%

\bibitem{Kubarovsky:2003fi}
V.~Kubarovsky {\it et al.}  [CLAS Coll.],
%``Observation of an exotic baryon with S = +1 in photoproduction from the
% proton,''
[Phys.\ Rev.\ Lett.\  {\bf 92}, 032001 (2004)]
Erratum -- ibid.\  {\bf 92}, 049902 (2004),
hep-ex/0311046.
%%CITATION = HEP-EX 0311046;%%

\bibitem{Togoo}
R.~Togoo {\it et al.}, Proc. Mongolian Acad. Sci., {\bf 4} (2003) 2.
% Carbon on Carbon --> p K_s, m_Theta=1532+-6

%\cite{Airapetian:2003ri}
\bibitem{Airapetian:2003ri}
A.~Airapetian {\it et al.}  [HERMES Coll.],
%``Evidence for a narrow $|$S$|$ = 1 baryon state at a mass of 1528-MeV in
%quasi-real photoproduction,''
Phys.\ Lett.\ B {\bf 585}, 213 (2004)
hep-ex/0312044.
%%CITATION = HEP-EX 0312044;%%

%\cite{Aleev:2004sa}
\bibitem{Aleev:2004sa}
A.~Aleev {\it et al.}  [SVD Coll.],
%``Observation of narrow baryon resonance decaying into p K0(S) in p A
%interactions at 70-GeV/c with SVD-2 setup,''
hep-ex/0401024.
%%CITATION = HEP-EX 0401024;%%

%\cite{Abdel-Bary:2004ts}
\bibitem{Abdel-Bary:2004ts}
M.~Abdel-Bary {\it et al.}  [COSY-TOF Coll.],
%``Evidence for a narrow resonance at 1530-MeV/c**2 in the K0 p system of
%the
%reaction p p $\to$ Sigma+ K0 p from the COSY-TOF experiment,''
hep-ex/0403011.
%%CITATION = HEP-EX 0403011;%%

%\cite{Aslanyan:2004gs}
\bibitem{Aslanyan:2004gs}
P.~Z.~Aslanyan, V.~N.~Emelyanenko and G.~G.~Rikhkvitzkaya,
%``Observation of S = +1 narrow resonances in the system K0(S) p from p +
%C-3_H-8 collision at 10-GeV/c,''
hep-ex/0403044.
%%CITATION = HEP-EX 0403044;%%

\bibitem{Chekanov:2004kn}
S.~Chekanov {\it et al.}  [ZEUS Coll.],
%``Evidence for a narrow baryonic state decaying to K0(S) p and K0(S)
%anti-p in
%deep inelastic scattering at HERA,''
hep-ex/0403051.
%%CITATION = HEP-EX 0403051;%%

\bibitem{NakanoNSTAR2004}
T. Nakano, talk at NSTAR 2004, March 24-27, Grenoble, France,
\hfill\break
{\tt http://lpsc.in2p3.fr/congres/nstar2004/talks/nakano.pdf}~.

\bibitem{Troyan:2004wp}
Y.~A.~Troyan {\sl et al.},
% A.~V.~Beljaev, A.~Y.~Troyan, E.~B.~Plekhanov,
%A.~P.~Jerusalimov, G.~B.~Piskaleva and S.~G.~Arakelian,
% ``The search and study of the baryonic resonances with the strangeness S = +1
%in the system of n K+ from the reaction n p $\to$ n p K+ K- at the momentum of
%incident neutrons P(n) = (5.20-GeV/c +- 0.12-GeV/c). (In Russian),''
hep-ex/0404003.
%%CITATION = HEP-EX 0404003;%%

\bibitem{ThetaWidth}
S. Nussinov,  hep-ph/0307357;
R.~W.~Gothe and S.~Nussinov,
 %``Are there non-strange low-lying pentaquarks and can we understand their
%width,''
hep-ph/0308230;
%%CITATION = HEP-PH 0308230;%%
%
R.~A.~Arndt, I.~I.~Strakovsky and R.~L.~Workman,
%``K+ nucleon scattering and exotic S = +1 baryons,''
Phys.\ Rev.\ C {\bf 68} (2003) 042201,
nucl-th/0308012 and nucl-th/0311030;
%
J. Haidenbauer and G. Krein,  hep-ph/0309243;
%
R.~N.~Cahn and G.~H.~Trilling,
%``Experimental Limits on the Width of the Reported Theta(1540)+,''
hep-ph/0311245;
%%CITATION = HEP-PH 0311245;%%
%
A.~Casher and S.~Nussinov,
%``The narrow Theta(1543). A QCD dilemma: Tube or not tube?,''
Phys.\ Lett.\ B {\bf 578} (2004) 124,
hep-ph/0309208.
%%CITATION = HEP-PH 0309208;%%
%--------------------------------------------------------------------
%% - negative results %%

\bibitem{BES}
J.~Z.~Bai {\it et al.}  [BES Collaboration],
%``Search for the pentaquark state in psi(2S) and J/psi decays to K0(S) p
%K-
%anti-n and K0(S) anti-p K+ n,''
hep-ex/0402012.
%%CITATION = HEP-EX 0402012;%%

\bibitem{HERA-B}
K.~T.~Knopfle, M.~Zavertyaev and T.~Zivko  [HERA-B Coll.],
%``Search for Theta+ and Xi(3/2)-- pentaquarks in HERA-B,''
hep-ex/0403020.
%%CITATION = HEP-EX 0403020;%%

\bibitem{PHENIX}
%\cite{Pinkenburg:2004ux}
C.~Pinkenburg [for the PHENIX Coll.],
%``Search for the anti-Theta- $\to$ K- anti-n with PHENIX,''
nucl-ex/0404001.
%%CITATION = NUCL-EX 0404001;%%

\bibitem{ALEPH}
P. Hansen [for ALEPH Coll.], talk at DIS 2004,
\hfill\break
{\tt http://www.saske.sk/dis04/talks/C/hansen.pdf}

\bibitem{DELPHI}
Throsten Wengler [reporting DELPHI Coll. results], talk at
Moriond '04 QCD,
\hfill\break
{\tt http://moriond.in2p3.fr/QCD/2004/WednesdayAfternoon/Wengler.pdf}

%----------------------------------------------------------------------
\bibitem{jenmalt}
For a review of the considerable theoretical literature on pentaquark models
and an in-depth discussion, see
Byron K. Jennings and Kim Maltman, hep-ph/0308286.

\bibitem{BurkertNSTAR2004}
V. Burkert, talk at NSTAR 2004, March 24-27, Grenoble, France,
\hfill\break
{\tt http://lpsc.in2p3.fr/congres/nstar2004/talks/burkert\_2.pdf}~.

\bibitem{Landsberg:1999wn}
 L.~G.~Landsberg, Phys.Rept.320(1999) 223;
hep-ex/9910048.

\bibitem{NA49Theta}
D. Barna [presenting NA49 Coll. data], lecture at
3rd Budapest Winter School on
Heavy Ion Physics, Dec. 8-11,
\hfill\break
{\tt http://www.hef.kun.nl/{\URLtilde}novakt/school03/agenda/Barnatalk.pdf}

\bibitem{OlPenta}
M.~Karliner and H.~J.~Lipkin,
%``The constituent quark model revisited: Quark masses, new predictions for
%hadron masses and K N pentaquark,''
hep-ph/0307243.
%%CITATION = HEP-PH 0307243;%%

\bibitem{NewPenta}
M. Karliner and H.J. Lipkin,
Phys.\ Lett.\ B {\bf 575} (2003) 249.

\bibitem{rosner} Jonathan L. Rosner, hep-ph/0312269.

\bibitem{Browder:2004mp}
T.~E.~Browder, I.~R.~Klebanov and D.~R.~Marlow,
%``Prospects for pentaquark production at meson factories,''
hep-ph/0401115.
%%CITATION = HEP-PH 0401115;%%

\bibitem{Armstrong:2003zc}
S.~Armstrong, B.~Mellado and S.~L.~Wu,
%``Prospects for pentaquark searches in e+ e- annihilations and gamma gamma
%collisions,''
hep-ph/0312344.
%%CITATION = HEP-PH 0312344;%%

\bibitem{CLAS-Trento}
M. Battaglieri [presenting CLAS Coll. data], talk at
Pentaquark Workshop, Feb. 10-12, 2004, Trento, Italy,
\hfill\break
{\tt http://www.tp2.ruhr-uni-bochum.de/talks/trento04/battaglieri.pdf}\ .


%\cite{Karliner:2004qw}
\bibitem{Karliner:2004qw}
M.~Karliner and H.~J.~Lipkin,
%``The narrow width of the Theta+: A possible explanation,''
Phys.\ Lett.\ B {\bf 586}, 303 (2004)
hep-ph/0401072.
%%CITATION = HEP-PH 0401072;%%

\bibitem{Aktas:2004pq}
A.~Aktas {\it et al.}  [H1 Collaboration],
%``Measurement of anti-deuteron photoproduction and a search for heavy
%stable
%charged particles at HERA,''
hep-ex/0403056.
%%CITATION = HEP-EX 0403056;%%

\bibitem{Knowles:1997dk}
I.~G.~Knowles and G.~D.~Lafferty,
%``Hadronization in Z0 decay,''
J.\ Phys.\ G {\bf 23}, 731 (1997)
hep-ph/9705217.
%%CITATION = HEP-PH 9705217;%%

\bibitem{Gustafson:1993mm}
G.~Gustafson and J.~Hakkinen,
%``Deuteron production in e+ e- annihilation,''
Z.\ Phys.\ C {\bf 61}, 683 (1994).
%%CITATION = ZEPYA,C61,683;%%

\bibitem{Akers:1995az}
R.~Akers {\it et al.}  [OPAL Collaboration],
%``Search for heavy charged particles and for particles with anomalous
%charge in
%e+ e- collisions at LEP,''
Z.\ Phys.\ C {\bf 67}, 203 (1995).
%%CITATION = ZEPYA,C67,203;%%

\bibitem{SloanPC}
T. Sloan, private communication.

\bibitem{JW}
R.~L.~Jaffe and F.~Wilczek,
%``Diquarks and exotic spectroscopy,''
Phys.\ Rev.\ Lett.\  {\bf 91}, 232003 (2003),
hep-ph/0307341.
%%CITATION = HEP-PH 0307341;%%

\bibitem{clodudek}
J.~J.~Dudek and F.~E.~Close,
%``The J = 3/2 Theta* partner to the Theta(1540) baryon,''
Phys.\ Lett.\ B {\bf 583}, 278 (2004),
hep-ph/0311258.
%%CITATION = HEP-PH 0311258;%%

\bibitem{DPP}
D.~Diakonov, V.~Petrov and M.~V.~Polyakov,
Z.\ Phys.\ A {\bf 359} (1997) 305,
hep-ph/9703373.
%%CITATION = HEP-PH 9703373;%%

\bibitem{Kopel}
H.~Walliser and V.~B.~Kopeliovich, J.\ Exp.\ Theor.\ Phys.\  {\bf
97} (2003) 433 [Zh.\ Eksp.\ Teor.\ Fiz.\  {\bf 124} (2003) 483]
hep-ph/0304058.
%%CITATION = HEP-PH 0304058;%%

\bibitem{other27}
D.~Borisyuk, M.~Faber and A.~Kobushkin, hep-ph/0307370 and
%%CITATION = HEP-PH 0307370;%%
%``Exotic baryons from the chiral quark soliton model,''
hep-ph/0312213;
%%CITATION = HEP-PH 0312213;%%
%
\hfill\break
 B.~Wu and B.~Q.~Ma,
%``The 27-plet baryons from chiral soliton models,''
hep-ph/0312041 and
%``Pentaquark Theta* states in 27 baryon multiplet from chiral soliton
%model,''
hep-ph/0312326.
%%CITATION = HEP-PH 0312326;%%

\bibitem{Ellis:2004uz}
J.~R.~Ellis, M.~Karliner and M.~Prasza{\l}owicz,
%``Chiral-soliton predictions for exotic baryons,''
hep-ph/0401127.
%%CITATION = HEP-PH 0401127;%%

\bibitem{JW2}
R.~Jaffe and F.~Wilczek,
%``Systematics of exotic cascade decays,''
hep-ph/0312369.
%%CITATION = HEP-PH 0312369;%%

\bibitem{NA49}
C.~Alt {\it et al.}  [NA49 Collaboration],
hep-ex/0310014.
%%CITATION = HEP-EX 0310014;%%

\bibitem{Fischer:2004qb}
H.~G.~Fischer and S.~Wenig,
%``Are there S=-2 Pentaquarks?,''
hep-ex/0401014.
%%CITATION = HEP-EX 0401014;%%

\bibitem{WA89}
J. Pochodzalla [presenting WA89 Coll. data], talk at the 2nd Panda Workshop,
Frascati, 3/18-19,2004,
%"Pentaquarks - facts and mysteries",
\hfill\break
{\tt
www.lnf.infn.it/conference/2004/Panda/Frascati2004\_final\_pochodzalla.pdf}~.

\bibitem{CDF}
I. Gorelov [presenting CDF Coll. data],
talk at DIS-2004, \v Strbsk\'e Pleso, Slovakia, 14-18 April
2004,
%\hfill\break
{\tt http://www.saske.sk/dis04/talks/C/gorelov.pdf}\ .

\bibitem{ChekanovDIS2004}
S. Chekanov [presenting ZEUS Coll. data], talk at DIS-2004, \v
Strbsk\'e Pleso, Slovakia, 14-18 April 2004,
%\hfill\break
{\tt http://www.saske.sk/dis04/talks/C/chekanov.pdf}\ .



\bibitem{varxistar}
M.~Karliner and H.~J.~Lipkin,
% A Mass Inequality for the $\Xi^*$ and $\Theta^+$ Pentaquarks,
hep-ph/0402008.

\bibitem{Karliner:2003si}
M.~Karliner and H.~J.~Lipkin,
%``The anticharmed exotic baryon Theta/c and its relatives,''
hep-ph/0307343.
%%CITATION = HEP-PH 0307343;%%

\bibitem{H1_Thetac}
A.~Aktas {\it et al.}  [H1 Collaboration],
%``Evidence for a narrow anti-charmed baryon state,''
hep-ex/0403017.
%%CITATION = HEP-EX 0403017;%%

\bibitem{ZEUS_Thetac}
L. Gladilin [on behalf of ZEUS Coll.], DESY seminar 12/3/04,
\hfill\break
{\tt http://webcast.desy.de/Pentaquark120304.htm}~;
\hfill\break
U. Karshon [on behalf of ZEUS Coll.], talk at DIS 2004,
\v Strbsk\'e Pleso, Slovakia, 14-18 April, \
%\hfill\break
{\tt http://www.saske.sk/dis04/talks/D/karshon.ps.gz}~.

\bibitem{FOCUS}
{\em ``A Search for a State which Decays to a Charged $D^*$
and Proton at FOCUS"},
{\tt  http://www-focus.fnal.gov/penta/penta\_charm.html}~.

\bibitem{AS}
Ya.I. Azimov and I.I. Strakovsky,
%Resonances, and mechanisms of Theta-production
hep-ph/0406312.



\end{thebibliography}
\end{document}